# Combining Deep Learning with Physics Based Features in Explosion-Earthquake Discrimination

**Qingkai Kong[1], Ruijia Wang[2,3], William R. Walter[1], Moira Pyle[1], Keith Koper[4], Brandon Schmandt[2]**

[1]Lawrence Livermore National Laboratory

[2]University of New Mexico

[3]Southern University of Science and Technology

[4]University of Utah

Corresponding author: Qingkai Kong (kong11@llnl.gov)

**Key Points:**

- Discrimination of earthquakes and explosions can be enhanced by combining physics-based features with those derived from machine learning

- Visualizing which parts of the input the deep learning model relies on can provide more insight into the processes underlying decisions

- The deep learning model focuses on different frequency bands for P and S waves to make the decision

**Abstract**

This paper combines the power of deep-learning with the generalizability of physics-based features, to present an advanced method for seismic discrimination between earthquakes and explosions. The proposed method contains two branches: a deep learning branch operating directly on seismic waveforms or spectrograms, and a second branch operating on physics-based parametric features. These features are high-frequency P/S amplitude ratios and the difference between local magnitude ($M_L$) and coda duration magnitude ($M_C$). The combination achieves better generalization performance when applied to new regions than models that are developed solely with deep learning. We also examined which parts of the waveform data dominate deep learning decisions (i.e., via Grad-CAM). Such visualization provides a window into the black-box nature of the machine-learning models and offers new insight into how the deep learning derived models use data to make the decisions.

**Plain Language Summary**

This paper presents a new method to distinguish earthquakes from explosions using seismic data. The method combines features implicitly defined by a deep learning algorithm with features explicitly defined from physical models of seismic sources and elastic wave propagation. The combination of these two types of features makes our method perform better on new data sets. By visualizing the performance of our combined model, we gain insight into what the deep learning model relies on to make the decisions.

**1 Introduction**

From creating catalogs of tectonic-only events for seismic hazard, to monitoring for nuclear explosions, discrimination between explosions and earthquakes remains an important task in seismology. Event characteristics such as focal depth, first motion polarity, efficiency of generating shear waves, have been found useful in distinguishing explosions from natural earthquakes for moderate size seismic events with magnitude larger than ~M3.5 (Bowers & Selby, 2009; National Research Council, 2012). More recently, with an interest in lowering monitoring thresholds, focus has turned to identifying smaller events that are well-recorded only at local distances (less than 250 km). For example, O'Rourke et al., (2016); Pyle & Walter, (2019, 2021) and Wang et al. (2020) showed that high-frequency P/S amplitude ratios can potentially be used for small-magnitude seismic discrimination by averaging over many stations at local distances. Furthermore, a recently proposed depth discriminant - the difference between local magnitude ($M_L$) and coda magnitude ($M_C$) - also shows the ability to separate explosions from deeper, naturally occurring earthquakes (Koper et al., 2021; Voyles et al., 2020). These physics-based discriminants provide a good understanding of the different characteristics between the two types of sources, and generally work very well in different regions and studies.

On the other hand, recent successful applications of machine learning to various areas in seismology (Bergen et al., 2019; Karpatne et al., 2019; Kong et al., 2019) suggest that a data-driven approach might be suitable for source classification problems. There are studies using machine learning models with manually selected features for source type discrimination (Dowla et al., 1990; Kong et al., 2016; Mousavi et al., 2016; Orlic & Loncaric, 2010; Rabin et al., 2016; Tsvang et al., 1993). These studies are often based on scientist-chosen features and use a machine learning algorithm to find the best classification boundary that best separates the source types. For example, Dowla et al. (1990) obtained a 97% rate of correct discrimination using features extracted at different frequencies from Pn, Pg, and Lg spectra with an artificial neural network to separate earthquakes from historic nuclear explosions at the former Nevada Test Site recorded on broadband seismic stations operated by Lawrence Livermore National Laboratory. These approaches usually work well if the features selected are representative of the different event sources, whereas features that are beyond scientists' awareness may also be missed. There have also been several studies using recently developed deep learning based approaches to distinguish explosions from natural earthquakes (Kim et al., 2020; Kong et al., 2021; Linville et al., 2019; Magana-Zook & Ruppert, 2017; Tibi et al., 2019). Linville et al. (2019) used convolutional and recurrent neural networks with spectrograms from seismic sensors as the input to classify explosions and tectonic sources at local distances, with a result of 99% accuracy in terms of the source type discrimination. Although deep learning methods have the advantage of extracting useful patterns from explosion and earthquake waveforms automatically without knowing any of the physics, they have the limitation that the learned features may have no clear physical meaning, and therefore may not generalize well in new regions. Even more problematic, they may focus on features such as the event location or timing, as opposed to a truly discriminating feature of the waveform.

In this paper, we propose to combine deep learning and physics-based features in one model for single-station explosion discrimination. By incorporating the physics-based features, the combined deep learning model improves on the performance compared to the deep learning model alone, especially when it is applied in a new region, i.e., improved transportability of the model. Furthermore, to uncover some of the black-box nature of the machine learning model, we also use model visualization methods to understand what the model learns to make the decision in identifying source type.

## 2 Data

Four datasets with local seismic observations of single-fired underground chemical explosions and earthquakes are used in this study. We use only valid P/S ratio measurements with values between 0 to 100 and recorded at stations within 250 km, limiting data to local distances. These studies are described in detail in the following cited original references, so only overview information is given here (station and shots distribution maps for the four regions are given in Figure S1).

The Source Physics Experiment (SPE) Phase I, conducted between 2011 and 2016, consisted of a series of underground chemical high-explosive detonations in saturated granite of various sizes and depths at the Nevada National Security Site. Phase I consisted of five $M_L$ 1.2 – 2.1 borehole shots (Snelson et al., 2013) and 110 earthquakes occurred with $M_L$ 1.0 – 4.4. In total, we assembled 149 explosion 3-component records and 2,216 earthquake records.

The Bighorn Arch Seismic Experiment (BASE) (Worthington et al., 2016; Yeck et al., 2014) was conducted in 2010 to image the Bighorn Arch in Wyoming. 21 explosive sources ($M_L$ 0.7 – 1.7) and 19 earthquakes ($M_L$ 0.3 – 2.7) were collected in the dataset, translating to a total of 4,394 explosion and 3,297 earthquake 3-component records.

The Mount St. Helens magma imaging project (MSH) contains 23 explosive sources ($M_L$ 0.9 – 2.3) and 91 earthquakes ($M_L$ 1.5 – 3.3) located within 75 km of Mt. St. Helens during the 2014 – 2016 iMUSH project (Kiser et al., 2016; Ulberg et al., 2020; Wang et al., 2020). In total, this dataset contains 1,652 explosion and 26,751 earthquake 3-component records.

The Salton Seismic Imaging Project (SSIP) is an active source seismic survey conducted in 2011 to image crustal faults and constrain rifting processes beneath the Salton Sea (Fuis et al., 2017; Han et al., 2016), during which 41 shots ($M_L$ 0.6 – 2.1) were conducted. The USGS reported 76 events in this region ($M_L$ 1 – 3.6) during the same time of the survey, including 6 borehole shots mis-labeled as earthquakes, as detailed in Wang et al. (2021). In total, the SSIP dataset has 2,307 explosion and 4,047 earthquake 3-component records.

We conduct a number of pre-processing steps to unify the waveforms from different regions. We first remove the mean and trend from each of the waveforms and apply a taper with a Hanning window followed by a four-corner bandpass filter from 1 – 20 Hz. We then resample the filtered waveforms to 40 Hz. In the last step, each waveform is cut to 2,000 data points (50 s) with a random start window (0-5 s) before the origin time. For each earthquake record, we randomly select the start time five times, while for each of the explosion records, 21 randomly selected start times are sampled. This data augmentation technique leads to a dataset of 173,385 earthquake and 178,059 explosion records that are roughly comparable for training and testing purposes. The raw data and augmented data distance, magnitude, depth and P/S ratio distributions are shown in Figures S2 and S3.

**3 Methods**

3.1 Two-branch model

3.1.1 Model structures

The proposed deep learning model contains two branches (Figure 1) to take advantage of both deep learning and physics-based parameters. It is a single-station-based binary classification problem (i.e., earthquake versus explosion). As shown, the blue dotted box contains the deep learning branch, where we use a convolutional neural network (Goodfellow et al., 2016; LeCun et al., 2015) to extract the features automatically from the input 3-component waveforms. The features extracted through the CNN (Convolutional Neural Network) layers are flattened into 128 features by the FC (Fully Connected) layer. The Physics Parameters Branch measures physics-based parameters that can be fed into this network; in our case, we use the P/S ratio and/or $M_L - M_C$. Features are extracted from these physics parameters by a FC layer and thus combined with those from the deep-learning branch by a Feature Concatenation layer. The concatenated features are then passed through another FC layer before the model makes the decision. Dropout layers that have been added after the CNN layers (dropout rate=0.3) and the Feature Concatenation layer (dropout rate=0.5) serve as regularization to reduce overfitting. ReLU (Rectified Linear Unit) activation functions are used across the network, except for the last layer, where two neurons with softmax activation functions (Goodfellow et al., 2016) are used to estimate the probability of the waveforms being earthquake or explosions. More details of the structure are shown in Figure S4.

In order to evaluate the effect of adding the physics parameters branch, we compare the performance of the two-branch model with that from a single-branch model. Two single-branch models are used: (1) the deep learning (CNN) model, which is the blue dotted box, and (2) the physics parameter branch model, which is the green dotted box shown in Figure 1.

3.1.2 Training and testing

During all trainings, we use the SparseCategoricalCrossentropy (Goodfellow et al., 2016) as our loss function and Adam as the optimization method (Kingma & Ba, 2017). The learning rate is initialized as 0.001. We also adopt an early stopping mechanism to avoid overfitting, with a maximum of 1,000 training epochs. If the validation accuracy does not change for 30 epochs, we stop the training and select the last highest accuracy model for the best trained model.

We use ROC (Receiver Operating Characteristic) curves on the testing data to quantify model performance, which jointly consider the true positive and false positive rates (Fawcett, 2006). In our case, the explosions are positive cases and earthquakes are negative cases. The Area Under the Curve (AUC) is used to measure the quality of the ROC curve, with a value of 1 being the best. To determine how adding the physics parameter branch helps, we adopt two data splitting methods. The first method randomly divides the whole dataset into training and testing, which is the common approach that has been used in many studies. The second method reserves one region for testing and uses the other three study regions for training, as four datasets are involved in this study. The latter method allows for better evaluation of model transportability with or without the physics parameter branch.

3.1.3 P/S ratio and $M_L$-$M_C$

P/S ratios are calculated from 10 – 18 Hz filtered three-component waveforms using the corresponding phases shown in equation (1) in Wang et al. (2021). SNRs (Signal-to-Noise Ratio) for the three component waveforms are also computed using the predicted P windows and noise windows (i.e., 10 s before the P arrival) for quality control purposes, with a cutoff threshold of SNR > 2 to declare a valid P/S ratio of a given event-station pair. More details about the parameters and regional velocity models used are described in (Wang et al., 2021).

To demonstrate that more physics-based features can be added, the "Adding $M_L$-$M_C$" section in the supplementary material provides design and testing for using the P/S ratio and the depth discriminant $M_L$-$M_C$ to improve the results slightly. But due to the limited number of $M_L$-$M_C$ measurements from the quality control, we will focus on using P/S ratio in the following sections.

3.2 Understanding what the model learns

For the deep learning branch, the Grad-CAM (Gradient-weighted Class Activation Mapping) is applied to understand what features are important to the deep learning branch to make the decision (Selvaraju et al., 2017). The basic idea behind this method is the last convolutional layer before flattening extracts the feature maps that contain the important features that the model relies on to make the decision. By taking the gradients of the final class score with respect to the feature maps, they provide a good indication of the pixels that are important to the final decision. Some examples of the Grad-CAM outputs are shown in the Supplement, where we overlap the derived heatmap on the input waveforms (Figure S8 – S10). In order to understand the role of different frequency bands, a deep learning model with spectrograms as inputs is added into the deep learning branch. The model structure is shown in Figure S5 (performance is shown in Figure S6), with Grad-CAM examples shown in Figure S11 – S13. These heatmaps illustrate the importance of the different regions on the input image (time series or spectrogram) in influencing the final output to the target class.

For P/S ratios in the physics parameters branch, we simply use the error rates (percentage of the wrongly estimated waveforms divided by the total waveforms) versus the P/S ratio value bins to show what the model learns.

**4 Results**

4.1 Model performance

When the testing data are from the same regions as the training data, using a single deep learning branch model (WF) has very similar results compared to the two-branch models (WF + PS) that included the P/S ratio branch (Figure 2a): both reach an AUC of 0.99. In contrast, using the P/S ratio alone (PS) leads to AUC=0.872. Such a difference clearly shows the power of deep learning versus a single parameter-based classifier. However, the model performance is significantly decreased when apply to data from a different region. In Figure 2b, where the training and testing data come from different regions, both WF + PS and the WF model performance degrade significantly on the new data, while the PS only model has a small degradation in performance. Due to the addition of the P/S ratio, the WF + PS model performs better than either single-branch model, showing the benefit of combining the physics-based features and the deep learning extracted features.

We validate the above observations and ensure they are generic through four sets of tests, where all sets of the three regions are used for training and the remaining one is used for testing (Figure 2c – 2f). The two-branch models achieve the best performance across different regions. Comparing the WF + PS to the WF models, the gaps show the contributions from the physics-based features, i.e., P/S ratio. Even though the deep learning models have the reputation of automatically extracting useful features, the physics-based features still can provide extra information that are not fully captured by deep learning, especially for data from new regions that are not in the training data.

4.2 Understanding what the model learned

Grad-CAM is used for understanding the deep learning branch. The calculated Grad-CAM weights are normalized to the range of 0 – 1, with higher values indicating greater importance. In order to obtain further understanding, we aggregate the weights by grouping them into 20 km distance bins and taking the average of the Grad-CAM weights (Figure 3).

For the earthquake records (Figure 3 row a), the P and S waves, including the time between the two arrivals, exhibit higher weights than other parts of the waveform. Interestingly, the importance of the weights drops significantly right after the S wave and then slowly increases again for some of the later parts of the waveforms. At further distances (i.e. >100 km), the model relies more on the P coda and S wave energy, likely due to the low P wave amplitudes at these large distances. The Grad-CAM weight distribution for the explosion records (Figure 3 row b) suggests heavier reliance on the P waves at closer distances, likely due to deficient S wave generation. At farther distances, a small and gentle peak from the later phases slowly takes over as the highest peak.

The Grad-CAM on the spectrograms (Figure 3 rows c and d using model in Figure S5) show similar patterns of the aggregated weights across the time dimension at different distances. For earthquake records, the model focuses on the P and S waves with increasing emphasis on S and coda at farther distances, while for explosion records, it mostly focuses on the P wave and the S coda. These figures also show extra information about the frequency content the model finds important. First, the energy between the P and S arrivals on the spectrograms appears to be less important than the analogous weights in the time series. Second, the model focuses on different frequency bands for P and S. The average of the five most important energy bands for each phase are shown in Figures 4a and 4b across different distances. The model relies on high-frequency P wave and low-frequency S wave energy to make the decisions. An apparent decay of the dominant frequency with distance were observed for the earthquake records but not explosions (Figures 4a and 4b). We speculate such distance-frequency dependence is associated with earthquake magnitudes, as higher number of small earthquakes with high corner frequencies are recorded at closer distances. The magnitude distribution for the explosion and earthquake records (Figure S18 and S19) support our explanation: the magnitude of explosions remains comparable across all distances, while the earthquakes show a positive correlation.

The results for understanding the contributions from the P/S ratios in the physics parameters branch are shown in Figure 4c and 4d. The model learns that larger P/S values are associated with explosion records, which are reflected by the higher error rates for earthquakes with high P/S ratios and lower error rates for explosion records with similar ratios.

## 6 Discussion and Conclusions

By adding physics-based features to the explosion discrimination problem, the two-branch model we propose here shows promising results for improving deep learning model performance, especially when the model is applied to data from a region different than the training region. One possible reason is that deep learning models are highly optimized to the training data, so they perform best when the testing data are from the same distribution. On the other hand, the physics-based features provide a more generally applicable basis, even though they may not perform as well as deep learning features on the training data. In combination, adding in physical features advances the performance on the new testing data, i.e., provides features not captured fully by the deep learning model. The combination essentially forces the model to learn from these physics-based features. While this application was to explosion discrimination, we envision a similar approach of incorporating physics-based features could be extended to other problems (e.g., noise and seismic discriminant in earthquake early warning, landslides classification and so on). Additional work combining physics and data learning is an exciting direction for future work.

Adding physics-based features does sacrifice the automated nature of the deep learning, however, it leverages prior human knowledge about what features are important and broadly applicable. Furthermore, the analysis shown here acts as a proof of concept, demanding extra work to implement it to real-world applications. One potential challenge is that P/S ratios may be unavailable for some waveforms (e.g., due to SNR thresholds). In this case, one can just rely on the results from the single deep learning branch, which requires a separate deep learning model to be trained on the waveforms. Alternatively, we could add one extra binary feature in the physics-parameter branch for each waveform to indicate if a measurement is available. This binary feature would work like a mask; during the training steps, when there are no P/S ratio measurements, the model would learn to rely only on the features extracted from the deep learning branch.

The discrimination performance is also sensitive to training data selection. The machine learning model tends to perform really well on the same distribution data, especially when the same data source is split for training and testing. As shown in Figures 2a and 2b, the machine learning models evaluated on the same distribution data may only reveal a limited picture, and may eventually fail in new data cases. This is known as local generalization of the machine learning models, i.e., they only perform well near where we have data points in the training data. Therefore, using a dataset from a different region can help reveal the limitation of the trained machine learning models.

Understanding what the trained model has learned from the waveform data can provide us with insights into how the model will perform in the real-world applications. The learning of the model from the physics-parameter branch aligns with the human knowledge, i.e., that higher P/S ratios correspond to explosion records. The Grad-CAM method reveals the focus areas of the trained model on the time series and the spectrograms. The information gained from P and S waves echoes seismologists' experience and supports the effectiveness of the machine-learning model. In addition, parts of the coda waves also contribute to the discrimination, especially at farther distances. On the spectrograms, the model weight peaks at different frequencies for different phases and event types, encouraging further tests to utilize different or dynamic frequencies for P/S ratio calculations (e.g., Tibi, 2021). In addition, P coda seems to be important for the time series, but not for the spectrograms, and it is less clear which parts of the coda matter most. Here, we only focused on preliminary interpretation of the results, and plan further work to develop a more complete understanding.

With the proof of concept of combining physics and data learning demonstrated here, we hope the proposed methods can be applied on different problems to build more reliable and extensible applications to real world issues of importance.

## Acknowledgments

This study uses data from the Source Physics Experiments (SPE), and SPE would not have been possible without the support of many people from several organizations. The authors wish to express their gratitude to the National Nuclear Security Administration, Defense Nuclear Nonproliferation Research and Development (DNN R&D), and the SPE working group, a multi-institutional and interdisciplinary group of scientists and engineers. The views expressed in the article do not necessarily represent the views of the U.S. Department of Energy or the U.S. Government. This research was performed in part under the auspices of the U.S. Department of Energy by the LLNL under Contract Number DE-AC52-07NA27344. This is LLNL Contribution LLNL-JRNL-829223. This research was partially supported by the Air Force Research Lab under contract FA9453-21-2-0024. We also thank the researchers of the datasets used in this study, including Source Physics Experiments (SPE), Bighorn Arch Seismic Experiment (BASE), Salton Seismic Imaging Project (SSIP), and imaging magma under the St. Helens magma (iMush; i.e., MSH). We thank the IRIS DMC (https://www.iris.edu/hq/) for hosting seismic data for research. The IRIS DMC is supported by the National Science Foundation under Cooperative Support Agreement EAR-1851048. The websites for these datasets are: SPE: **https://doi.org/10.7914/SN/CI**, **"IM"(no DOI), "LB"(no DOI),** **https://doi.org/10.7914/SN/NN**, **https://doi.org/10.7914/SN/SN**, **https://doi.org/10.7914/SN/TA**, **https://doi.org/10.7914/SN/US**, **https://doi.org/10.7914/SN/UU**, **https://doi.org/10.7914/SN/XE_2012**; BASE: **https://doi.org/10.7914/SN/IW**, **https://doi.org/10.7914/SN/IU**, **https://doi.org/10.7914/SN/MB**, **"PB"(no DOI), "RE"(no DOI),** **https://doi.org/10.7914/SN/TA**, **https://doi.org/10.7914/SN/US**, **https://doi.org/10.7914/SN/WY**, **https://doi.org/10.7914/SN/XV_2009**, **"Z2"(no DOI),** **https://doi.org/10.7914/SN/ZH_2010**; MSH: **SN/CC"** **https://doi.org/10.7914/SN/CC**,

https://doi.org/10.7914/SN/XD_2014, https://doi.org/10.7914/SN/UW; and SSIP: https://doi.org/10.7914/SN/AZ, https://doi.org/10.7914/SN/CI, https://doi.org/10.7914/SN/XD_2011. We thank useful discussions from Stephen Myers, Giselle Fernández-Godino, and Donald Lucas at the Lawrence Livermore National Laboratory. All the analysis is done in Python and the deep learning framework used here is TensorFlow (Abadi et al., 2016), and seismological related analysis used Obspy (Beyreuther et al., 2010; Krischer et al., 2015), we thank the awesome Python communities to make everything openly available.## References

Abadi, M., Barham, P., Chen, J., Chen, Z., Davis, A., Dean, J., et al. (2016). TensorFlow: A system for large-scale machine learning. In *12th USENIX Symposium on Operating Systems Design and Implementation (OSDI 16)* (pp. 265–283). Retrieved from https://www.usenix.org/system/files/conference/osdi16/osdi16-abadi.pdf

Bergen, K. J., Johnson, P. A., Hoop, M. V. de, & Beroza, G. C. (2019). Machine learning for data-driven discovery in solid Earth geoscience. *Science*, *363*(6433). https://doi.org/10.1126/science.aau0323

Beyreuther, M., Barsch, R., Krischer, L., Megies, T., Behr, Y., & Wassermann, J. (2010). ObsPy: A Python Toolbox for Seismology. *Seismological Research Letters*, *81*(3), 530–533. https://doi.org/10.1785/gssrl.81.3.530

Bowers, D., & Selby, N. D. (2009). Forensic Seismology and the Comprehensive Nuclear-Test-Ban Treaty. *Annual Review of Earth and Planetary Sciences*, *37*(1), 209–236. https://doi.org/10.1146/annurev.earth.36.031207.124143

Dowla, F. U., Taylor, S. R., & Anderson, R. W. (1990). Seismic discrimination with artificial neural networks: Preliminary results with regional spectral data. *Bulletin of the Seismological Society of America*, *80*(5), 1346–1373. https://doi.org/10.1785/BSSA080005134610

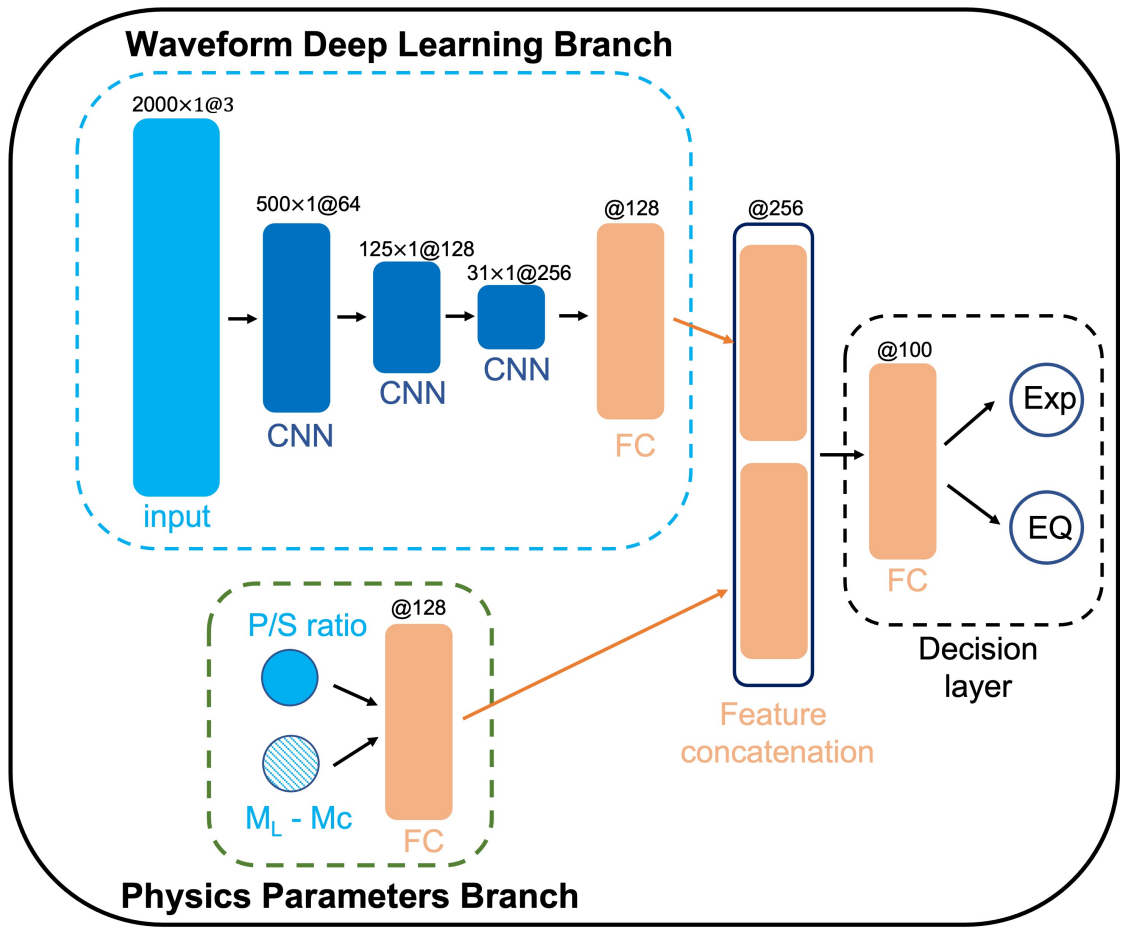

Figure 1. The two-branch proposed model. The blue dotted box is the deep learning branch which takes in waveforms as the input with convolutional neural network (CNN) layers as the hidden layers, and a fully connected layer (FC) to flatten out the features. The green dotted box contains the physics-based feature branch, which can take in P/S ratio measurements with or without the $M_L$-$M_C$ measurements (adding $M_L$-$M_C$ is shown in supplementary materials). Features from these two branches will be concatenated and pass another FC layer to make a decision. The small texts on top of the layer block represent the feature maps in CNN or number neurons in FC. 500x1@64 represents 64 feature maps with dimension 500x1.

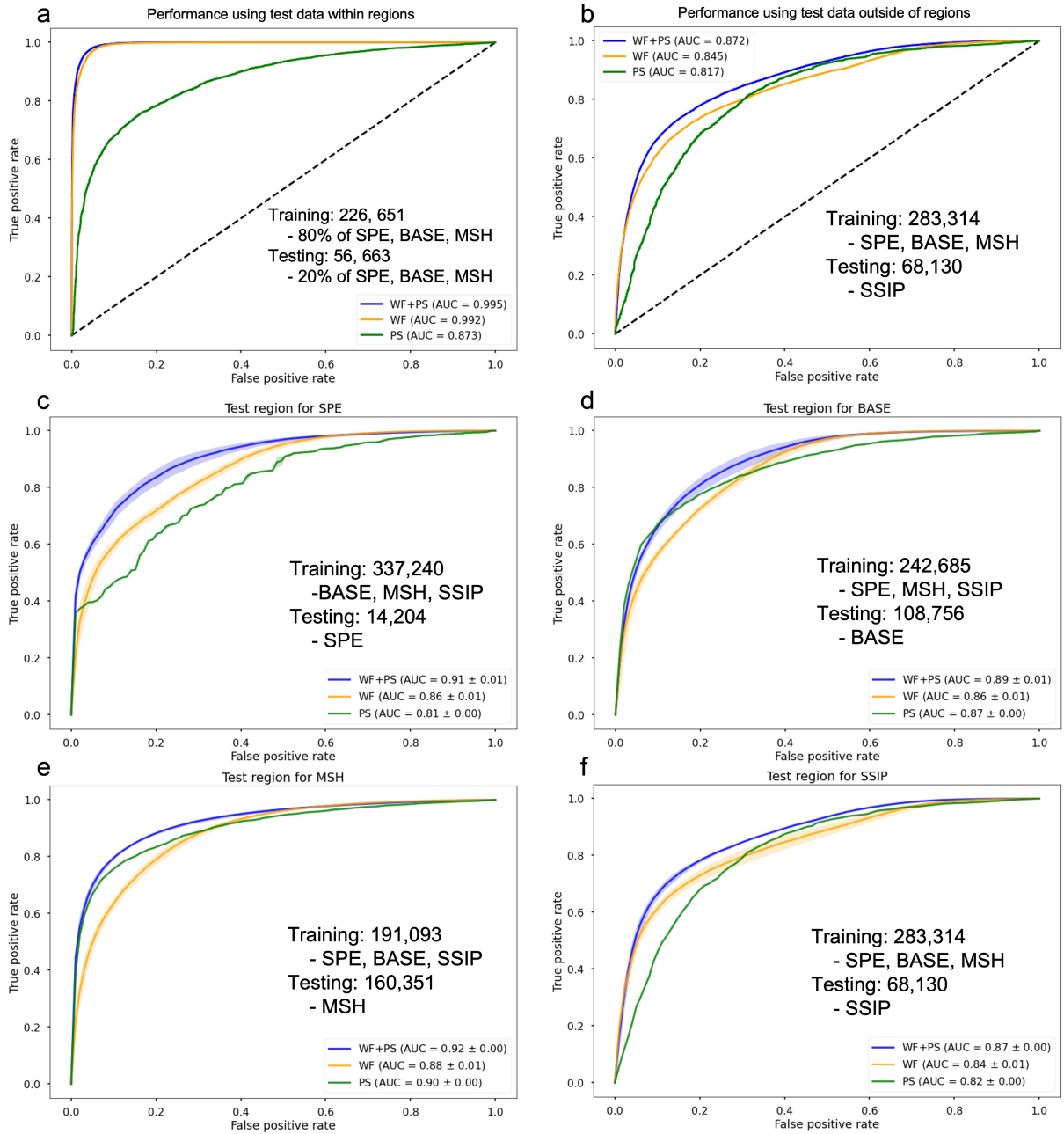

Figure 2. Classification performance metrics. (a) cases where testing data is from the same region as training data, i.e., SPE, BASE, and MSH (20% of the total data saved as testing data). (b) cases where testing data is from a different region, SSIP, rather than the three training regions (c – f) The ROC curve using training data from any of the three regions and testing on the new fourth region for five random initialization with mean (solid lines) and standard deviation (shaded areas). The blue curves show the designed model with deep learning and physics parameters branches. The orange and green curves are the model only with the deep learning or physics parameters branch. The AUC is shown in the legend. WF – Waveform, PS – P/S ratio.

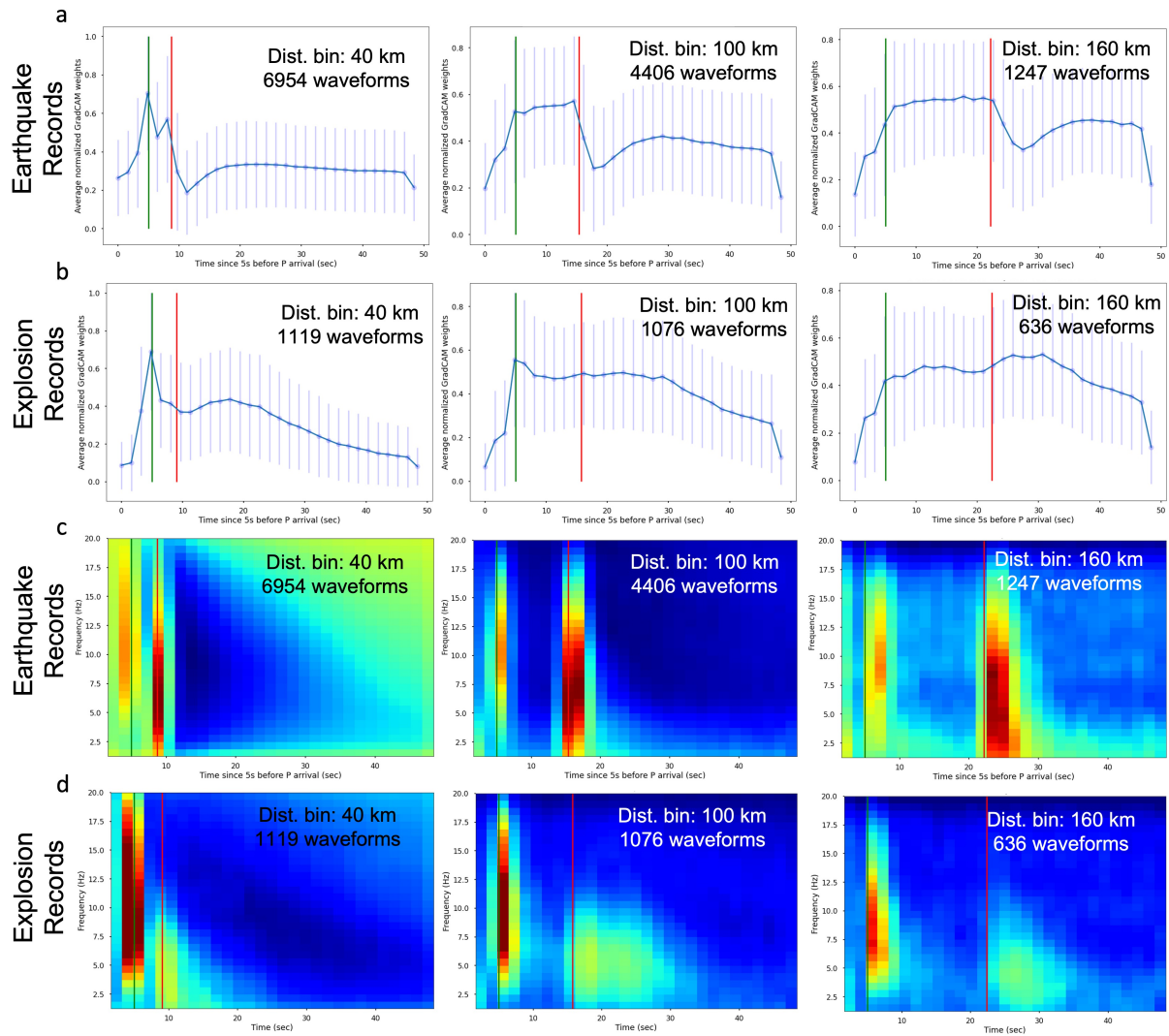

Figure 3. Average of the normalized Grad-CAM weights for the earthquake and explosion records across different distance bins (bin size 20 km). Rows a and b are weights on the time series. The blue thick lines are the average weights and the vertical thin blue lines are the standard deviation in the bins. The vertical green and red lines are the average of the estimated P and S arrivals using the same regional velocity models as in the calculation of the P/S ratios. For each panel, the horizontal axis is the time in seconds starting 5s before the arrival of P wave. The vertical axis is the normalized weight. The rows c and d are weights on the spectrograms. Color shows the weights, from blue (0) to red (clipped at 0.5). Vertical green and red lines are the average of the estimated P and S arrivals using the same regional velocity models as in the calculation of the P/S ratios. For other distance bins, please refer to Figures S14 – S17.

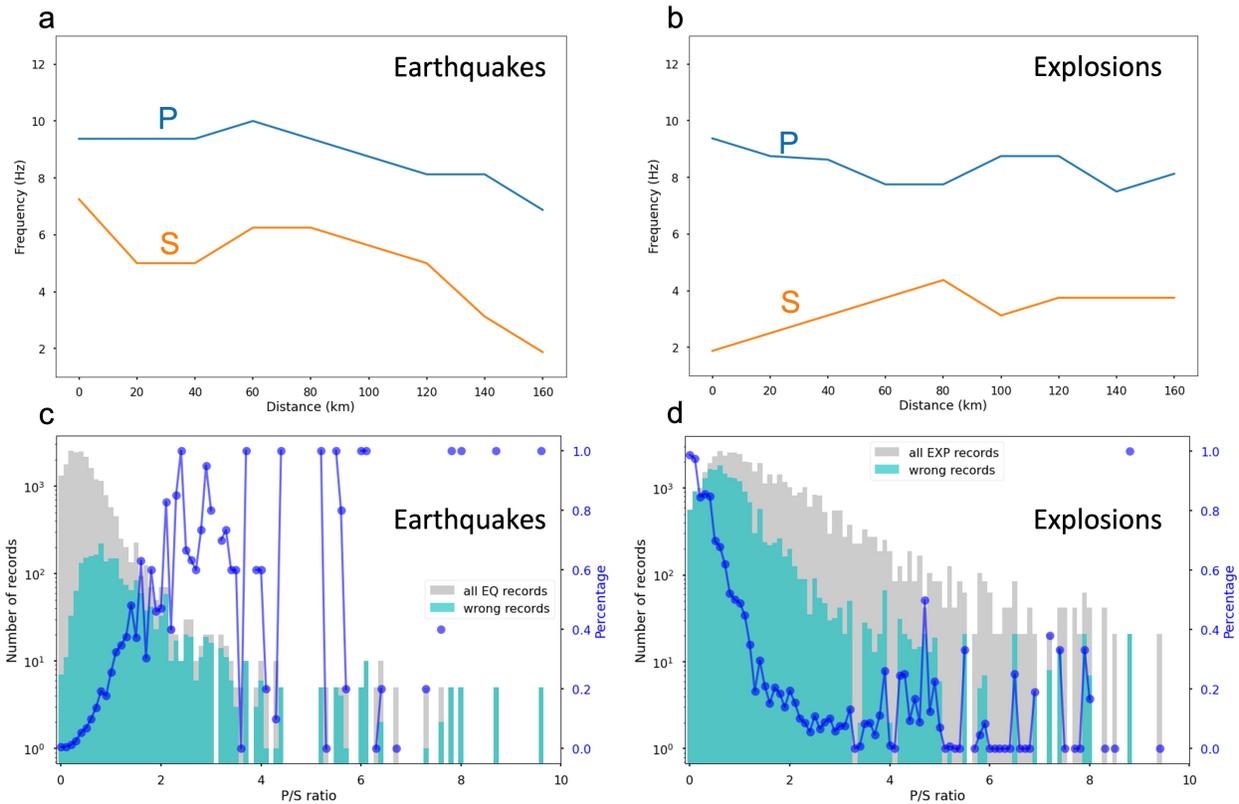

Figure 4. The average of the top 5 frequency bins across different distances corresponding to Figures S16 and S17 and Physics-based branch error visualization. (a) For the earthquake records. (b) For the explosion records. The blue lines are for the P wave while the orange lines are for the S wave. (c) P/S ratio error histogram for earthquake records with 0.1 step bins. The grey bars are all the earthquake records, and the cyan colored bars are the wrongly estimated records in each bin. The blue line with dots is the percentage calculated using the cyan bar over grey bar, i.e. the error rate. (d) P/S ratio error histogram for earthquake records with 0.1 step bins. Same as (c), but for the explosion records.

# Supplementary Materials for Combining Deep Learning with Physics Extracted Features in Explosion-Earthquake Discrimination

Qingkai Kong[1], Ruijia Wang[2,3], William R. Walter[1], Moira Pyle[1], Keith Koper[4], Brandon Schmandt[2]

[1]Lawrence Livermore National Laboratory

[2]University of New Mexico

[3]Southern University of Science and Technology

[4]University of Utah

Corresponding author: Qingkai Kong (kong11@llnl.gov)

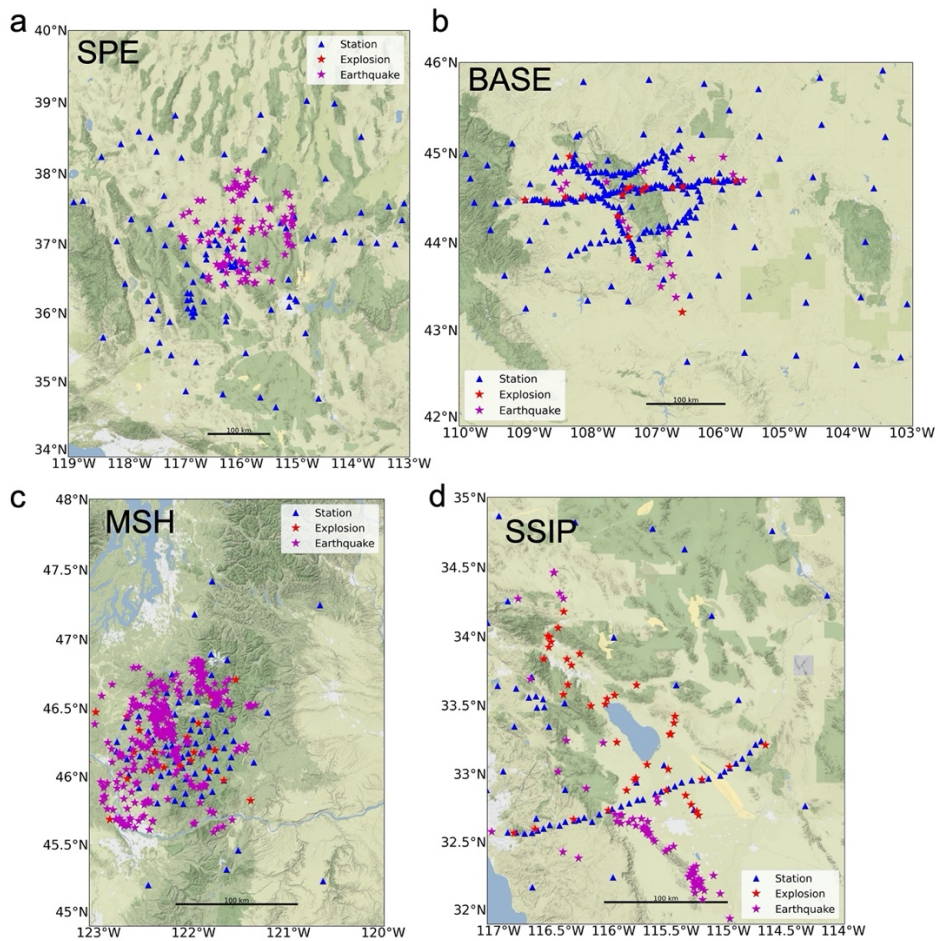

Figure S1. Events and station maps for the four experiments used in the study. The red and magenta stars are explosion and earthquake events, respectively. The blue triangles show the distribution of the stations. (a) Map for the Source Physics Experiment. (b) Map for the Bighorn Arch Seismic Experiment. (c) Map for the Mount St. Helens magma imaging experiment. (d) Map for the Salton Seismic Imaging Project.

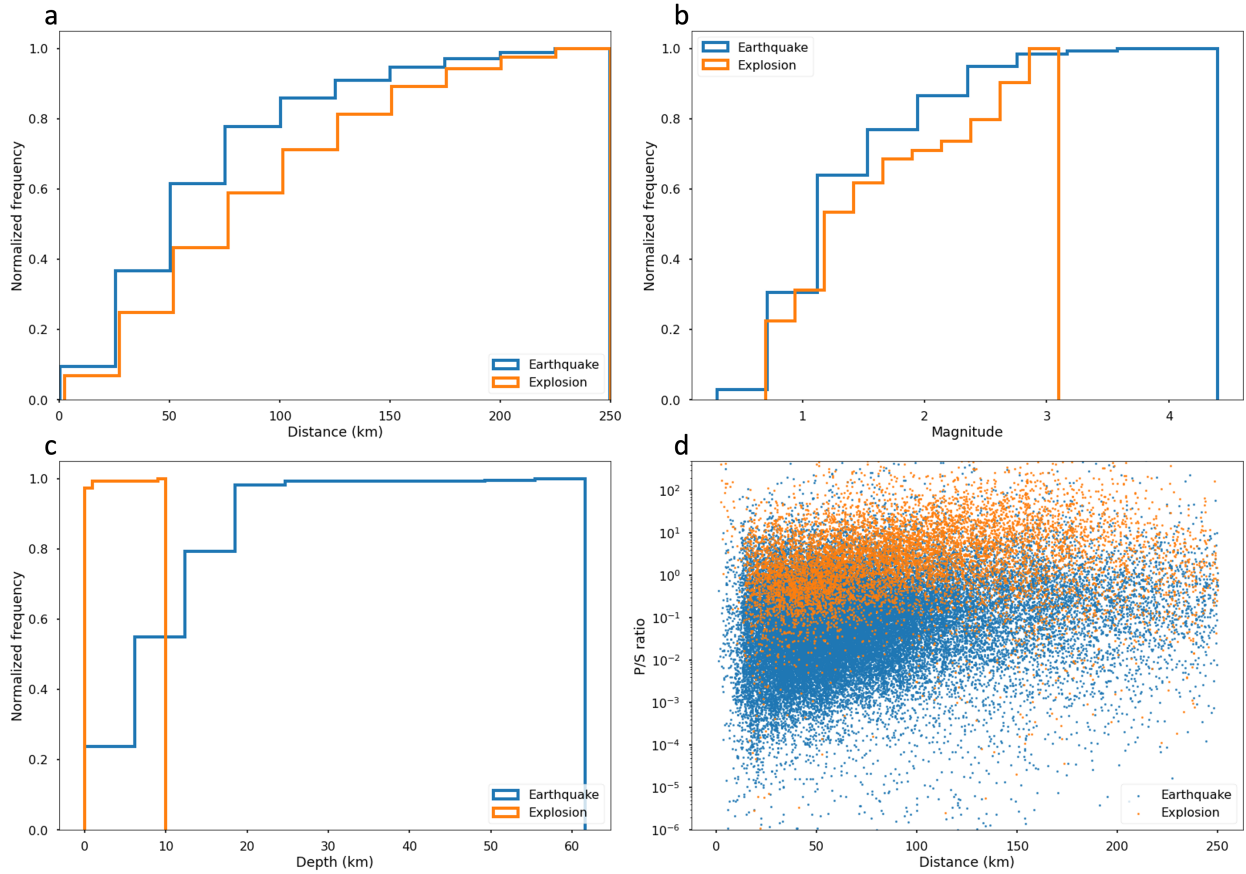

Figure S2. Raw data distribution from the four different regions, i.e., BASE, SPE, MSH and SSIP. (a) cumulative distance distribution. (b) cumulative magnitude distribution. (c) cumulative depth distribution. (d) scatter plot for P/S ratios over distance.

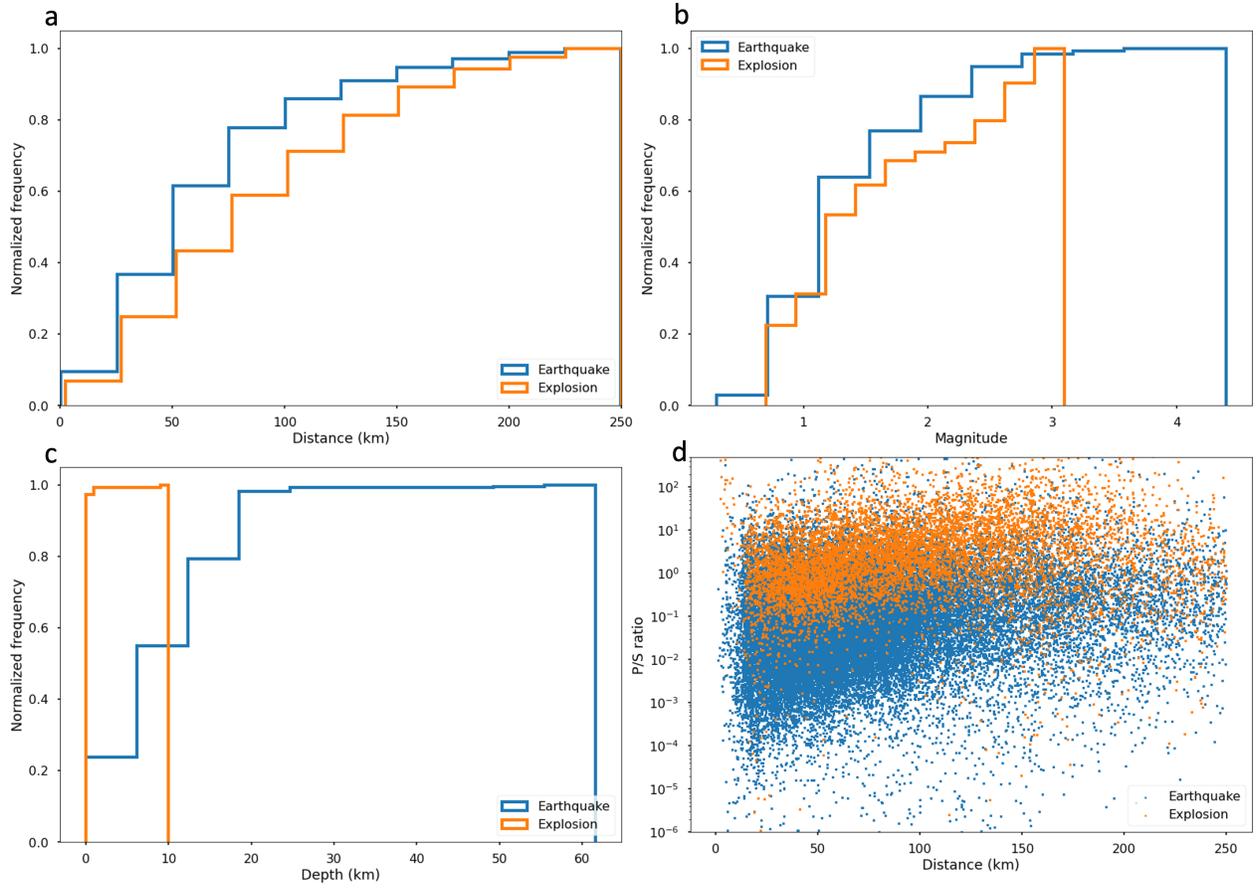

Figure S3. Augmented data distribution from the four different regions, i.e., BASE, SPE, MSH and SSIP. Panels are the same as Figure S2.

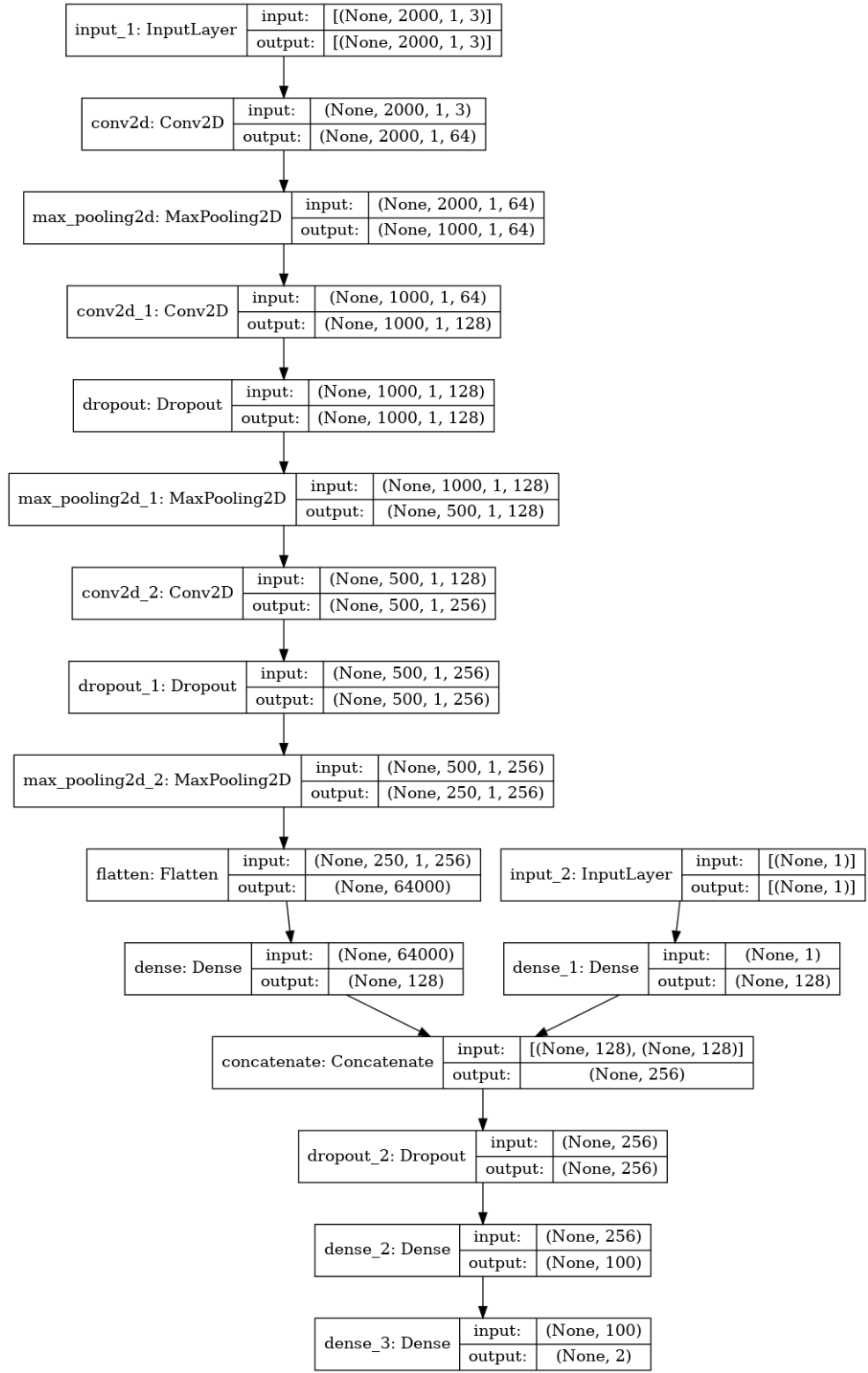

Figure S4. The corresponding model architecture from tensorflow for Figure 1 in the main text. This version shows more information about the structure, such as input/output dimension and the dropout layers.

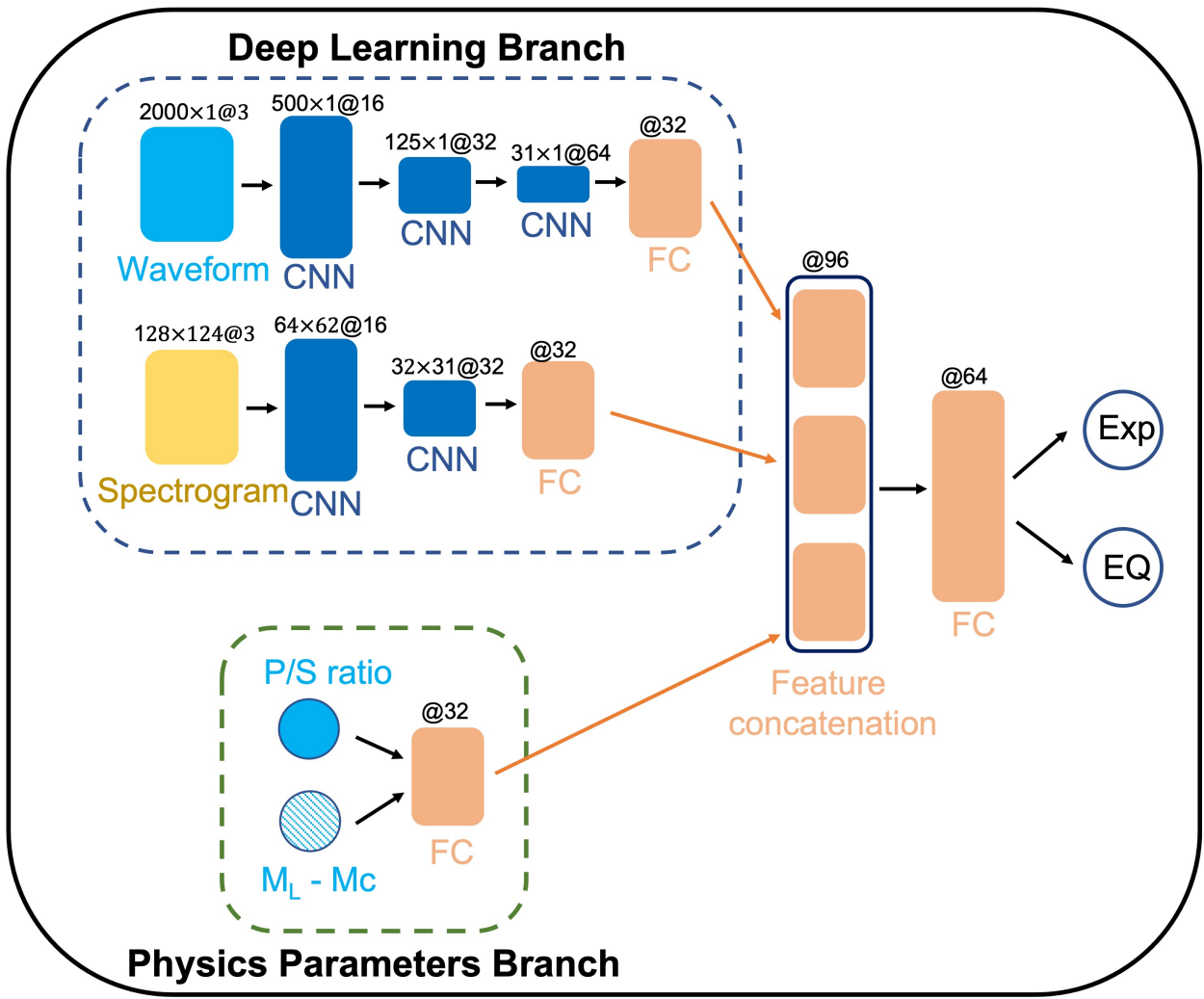

Figure S5. The model architecture for the proposed model with the additional sub-branch to take in spectrograms as the input in the deep learning branch.

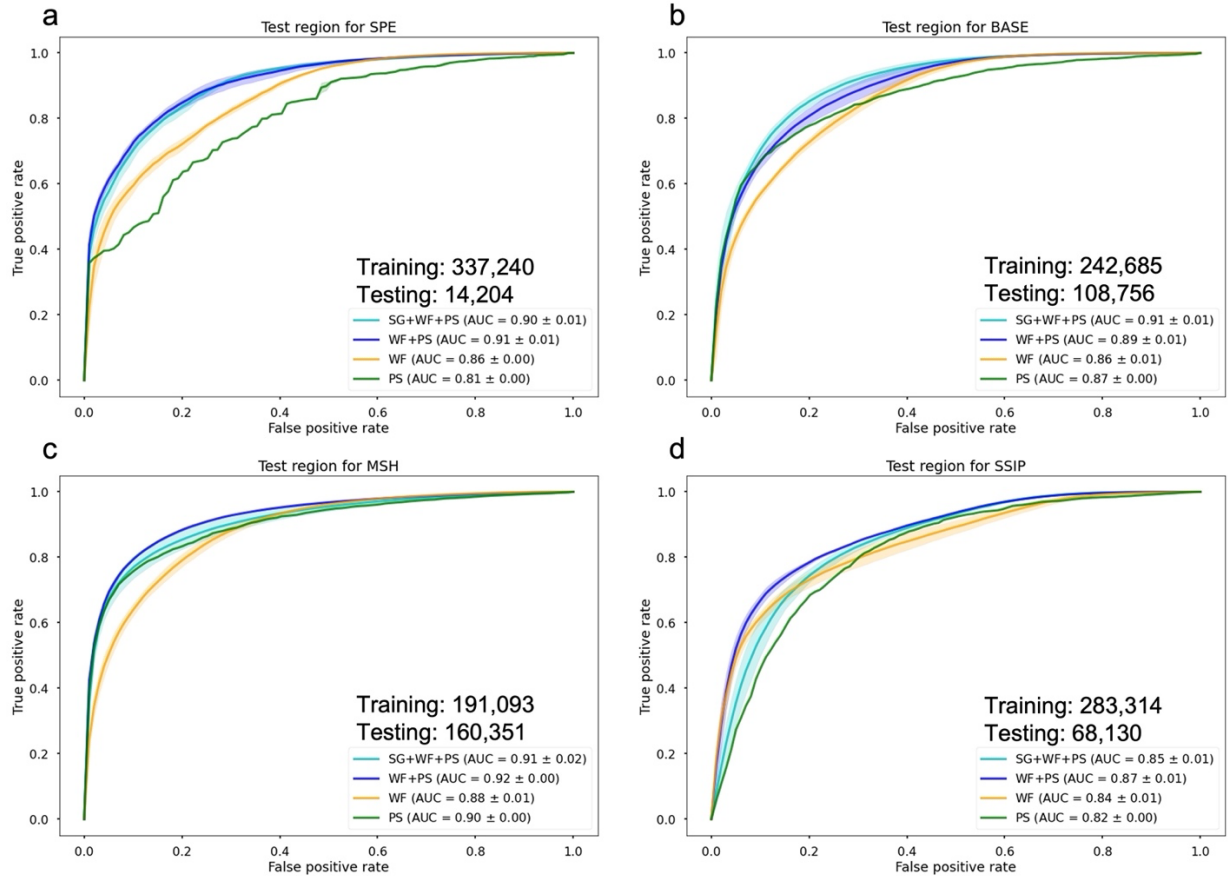

Figure S6. Test results corresponding to the model structure shown in Figure S5. Similar to Figures 2c – 2f, but added the performance of the model with the spectrogram. SG – Spectrogram input, WF – waveforms, PS – P/S ratios. Thus, SG + WF + PS is the two-branch model with the spectrogram, time series, and P/S ratios inputs.

**Adding $M_L$ - $M_C$**

$M_L$ and $M_C$ values are all calculated directly from the waveforms. For computing $M_L$, the HH and BH component waveforms are converted to a Wood-Anderson seismometer equivalent response. Amplitudes are measured without station corrections, and distance corrections are applied. $M_C$ is calculated from the measured coda duration using the equations described in Koper et al. (2021). Due to the quality control threshold (more details see Wang et al. 2021), only 59,623 waveforms have corresponding measured $M_L$-$M_C$ values.

As expected, the models combining the deep learning extracted features with these two physics-based features perform the best across different regions, though the differences from the models using only the P/S ratios are negligible. Although $M_L$-$M_C$ model alone shows the poorest performance compared to other models, adding these measurements to some regions improves the results (e.g., SPE and BASE). Note that due to quality control requirements for valid $M_L$- $M_C$ measurements and P/S ratios, the datasets used for the above tests are much smaller than those shown in Figure S6.

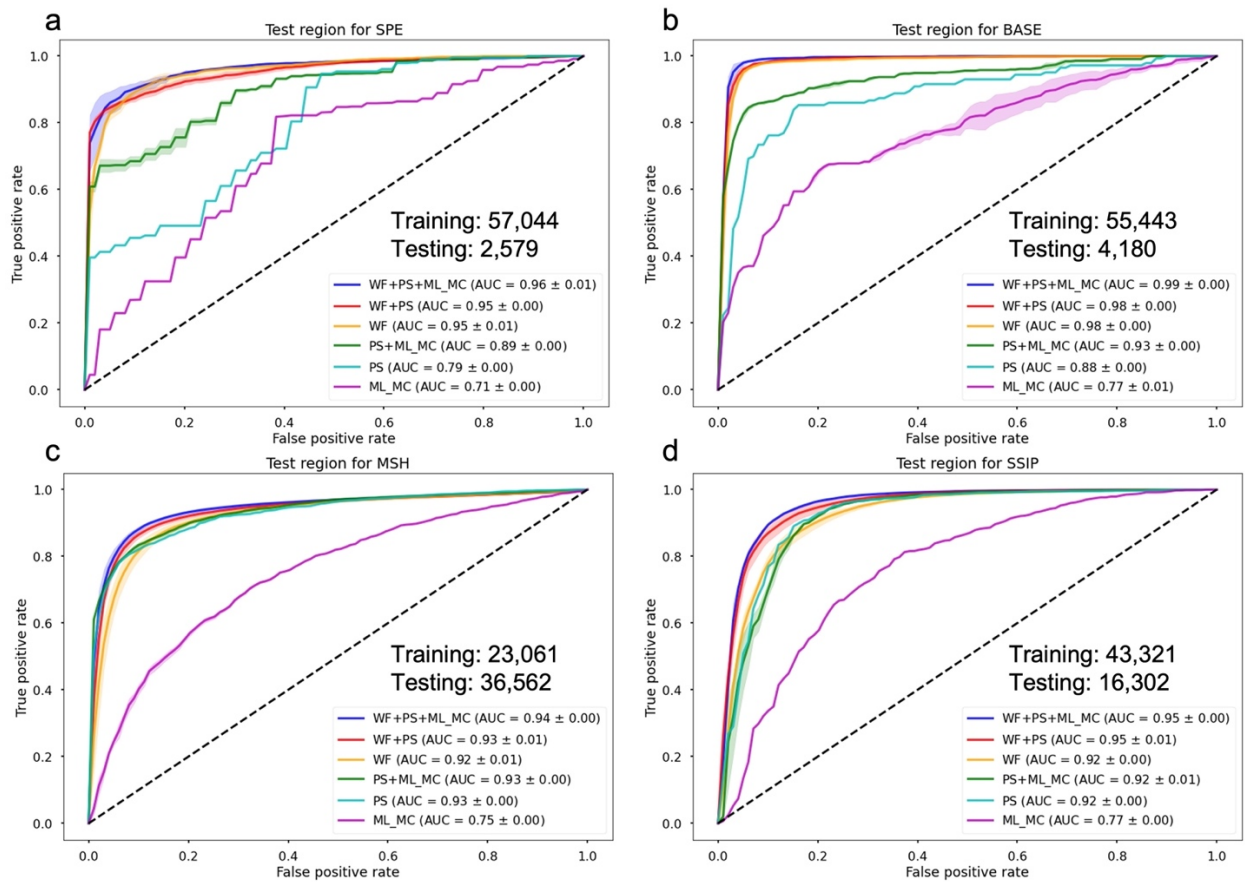

Figure S7. The ROC curve using training data from any of the three regions and testing on the new fourth region with $M_L$-$M_C$. WF + PS + ML_MC are the two-branch models with P/S ratios and $M_L$-$M_C$ values used in the physics-based branch, model structure is shown in Figure 1. PS+ML_MC are the single-branch physics-based models with P/S ratio and $M_L$-$M_C$ values as the inputs. The ML_MC are the models that only use $M_L$-$M_C$ values as the input.

**Grad-CAM**

Following the symbols used in Selvaraju et al. (2017), we can first take the gradient of the score $y^c$ for class c, with respect to the last CNN layer feature maps $A^k$, where k is the k-th feature map, and by globally averaging them across all the pixels, we can define the importance weights $\alpha^c_k$ for each feature map in the following equation

$$\alpha^c_k = \frac{1}{z}\sum_i^u \sum_j^v \frac{\partial y^c}{\partial A^k_{ij}} \qquad (1)$$

where u and v are the width and height of the feature map, and i and j are the (i, j) pixel on the feature map. The relative importance of each of the feature maps are given by these weights for the target class c before the model makes a decision. Then the class discriminative localization map can be calculated using equation

$$L^c_{Grad-CAM} = ReLU(\sum_k^n \alpha_k^c A^k) \quad (2)$$

where n is the total number of feature maps in the last convolutional layer, and ReLU (Rectified Linear Unit) is an operation that is the same as the activation functions used in the model; mathematically it is defined as max(0, x). The reason to apply a ReLU to the linear combination of the weighted feature maps is that only the features that have a positive influence on the target class are needed. The derived localization map is a non-negative weighted average of all the feature maps in the same dimension of the last feature map, and it is up-sampled to the input image resolution using bi-linear interpolation to form the final heatmap.

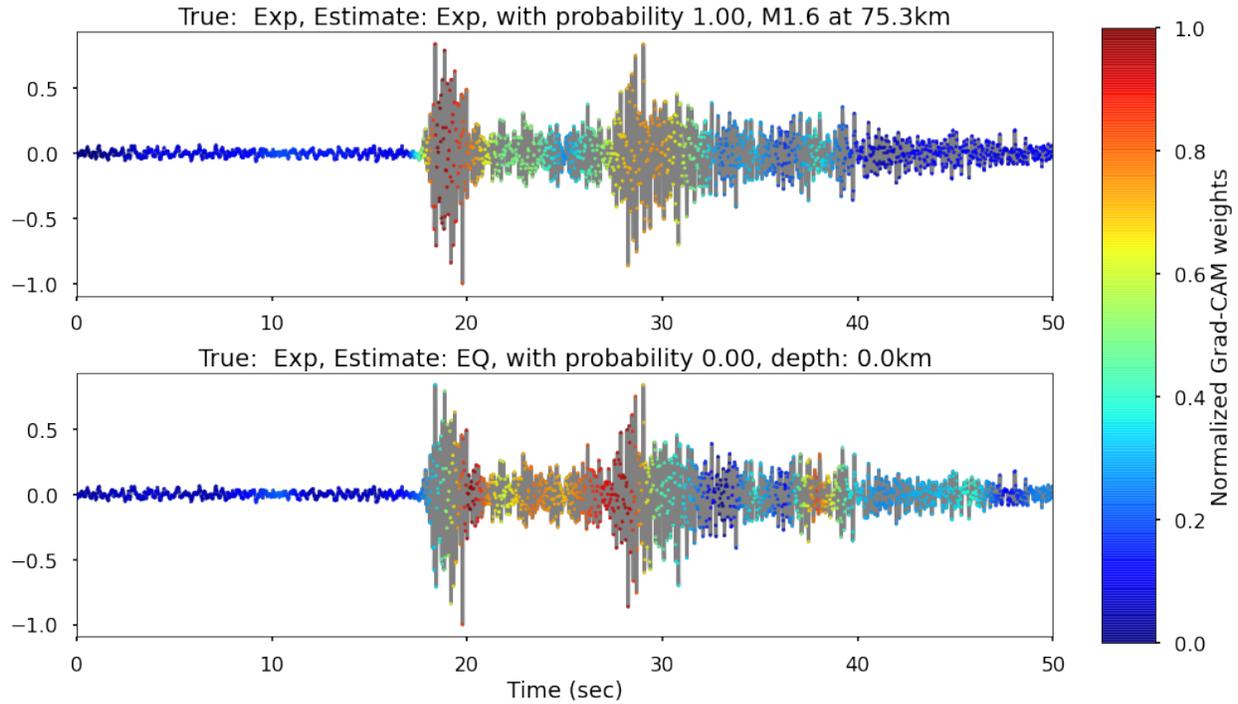

Figure S8. Grad-CAM on time series of a $M_L1.6$ explosion record recorded at 75.3 km. The titles show the estimation, ground truth, and the estimation probability. For example, the top row shows that the model estimates the input as an explosion with probability=1. The bottom row shows the earthquake estimation with probability 0.

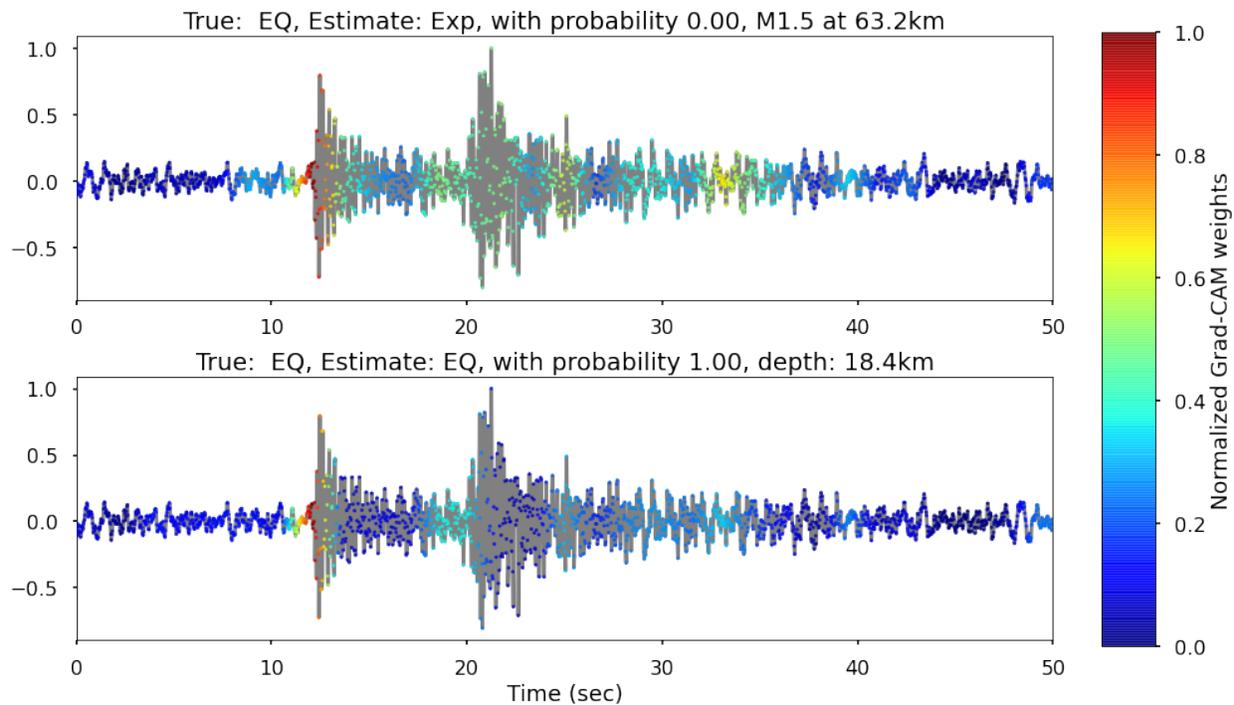

Figure S9. Grad-CAM on a time series recorded at 63.2 km for a $M_L$1.5 and 18.4 km depth event. Legend is the same as Figure S8.

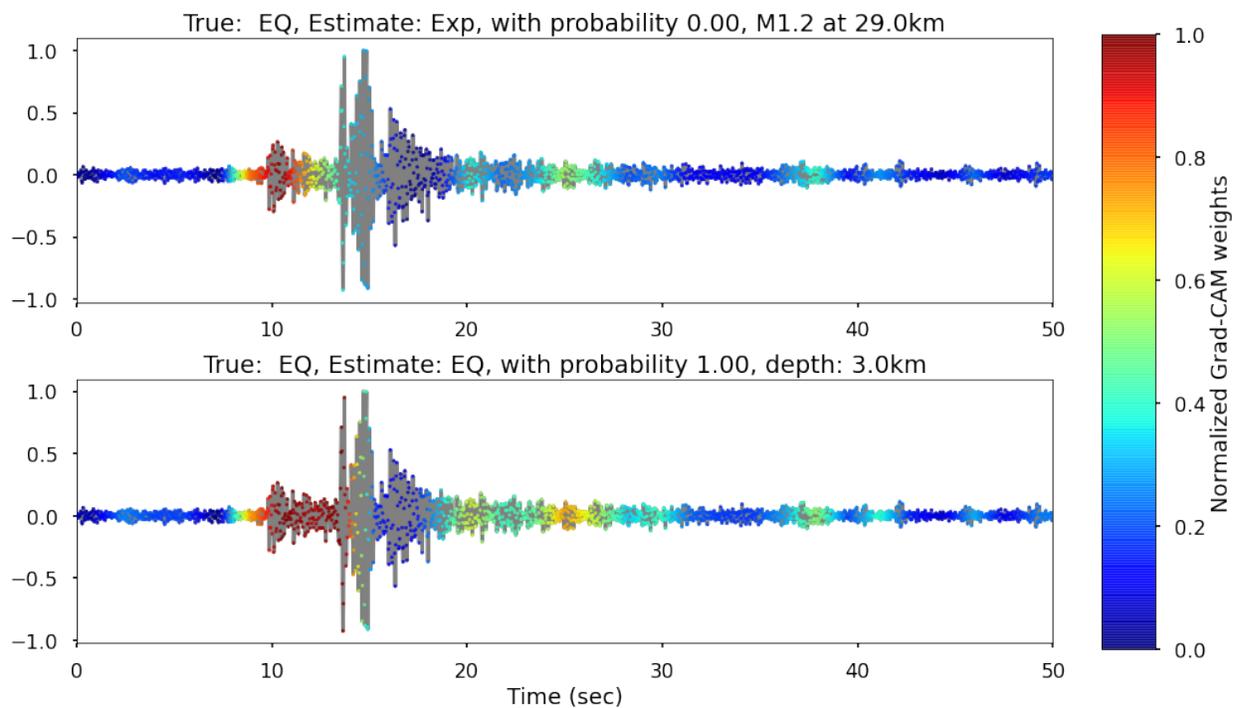

Figure S10. Grad-CAM on a time series recorded at 29.0 km for a $M_L$1.2 and 3.0 km depth event. Legend is the same as Figure S8.

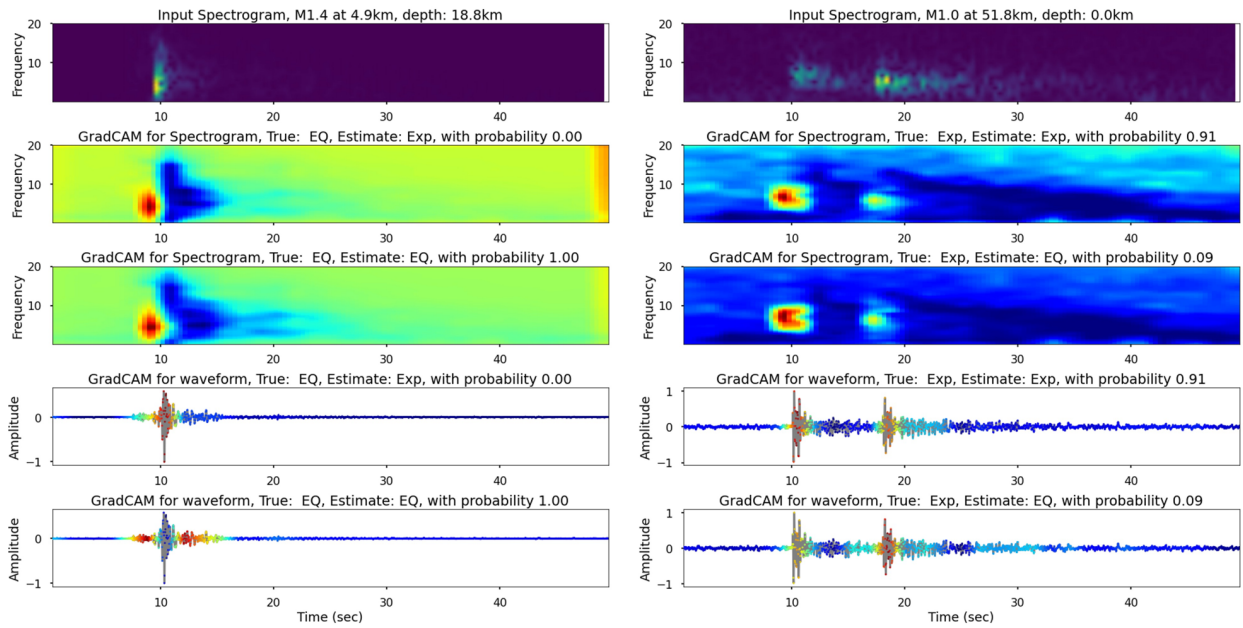

Figure S11. Grad-CAM heatmaps with the inputs. Two examples are shown here as two columns. For each column, the top row shows the spectrogram inputs for the model, with title indicating the magnitude, epicentral distance, and depth of the event, i.e., $M_L$1.4 earthquake with depth 18.8 km and $M_L$1.0 explosion with depth 0.0 km. The second and third rows show the Grad-CAM heatmap for the spectrogram input with warmer colors showing more important regions for the model to rely on to make the decision. The titles indicate the estimation, ground truth, and the estimation probability. For example, the second row shows the model estimates the input as an explosion with probability 0 for the left panel, and 1 for the right. The third row shows the earthquake estimation probability. The fourth and fifth rows are the same as the second and third rows but for the time series inputs.

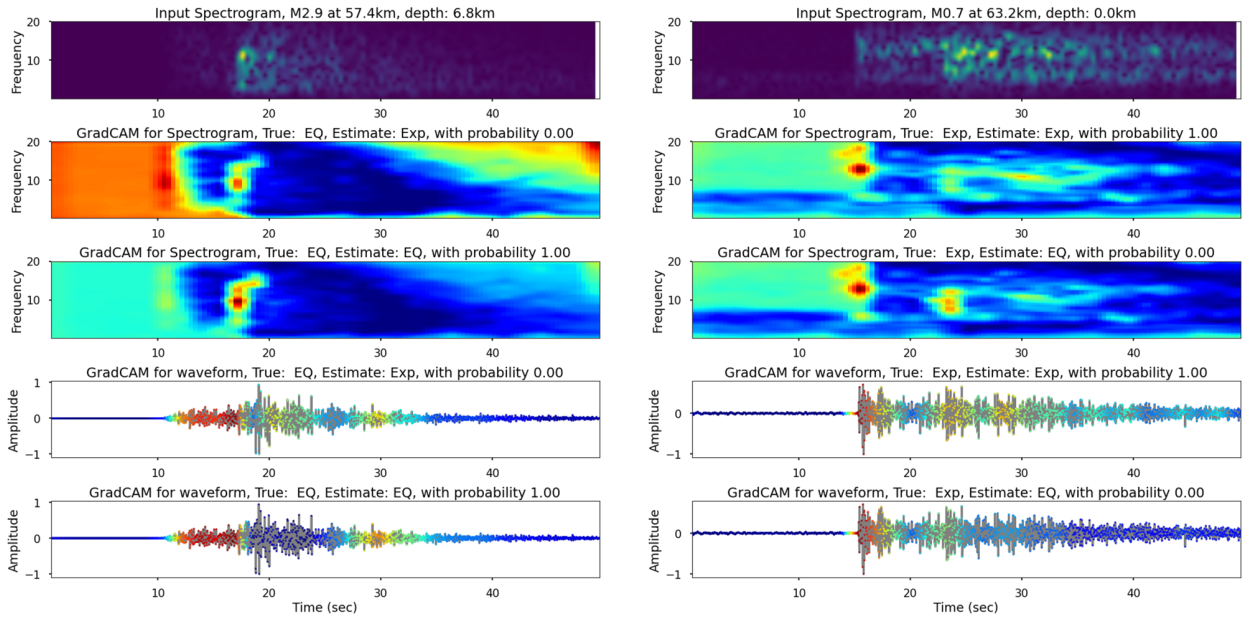

Figure S12. Same with S11 but for two different events, i.e., $M_L$2.9 earthquake with 6.8 km depth and $M_L$0.7 explosion with 0.0 km depth.

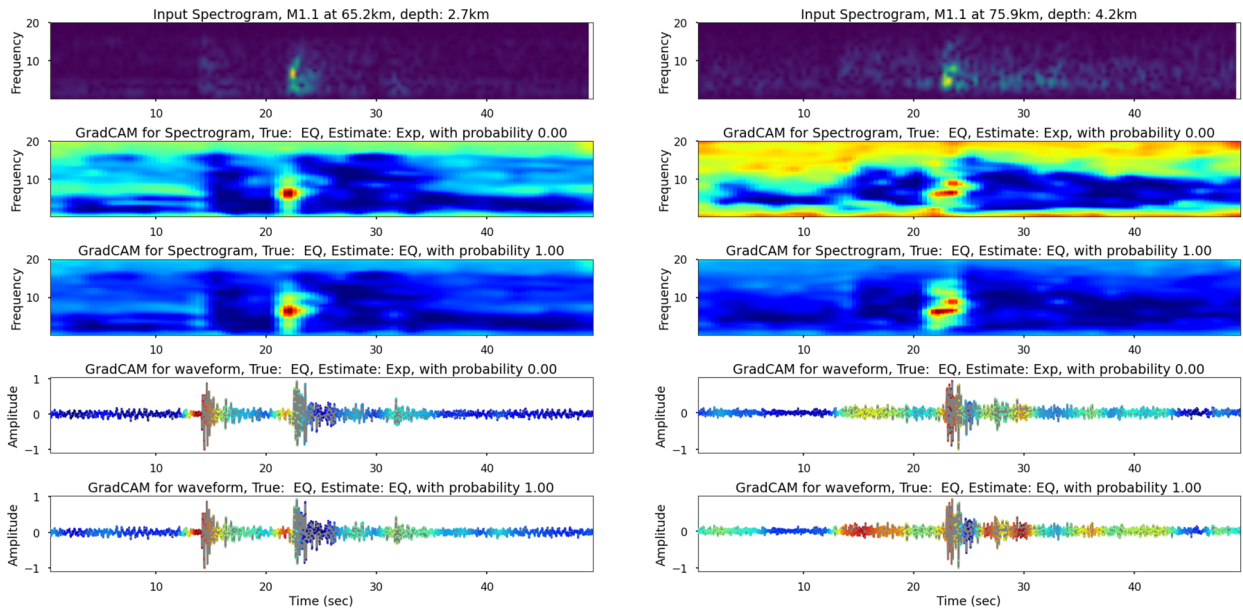

Figure S13 Same with S11 but for two different events, i.e., $M_L$1.1 earthquake with 2.7 km depth and $M_L$1.1 with 4.2 km depth.

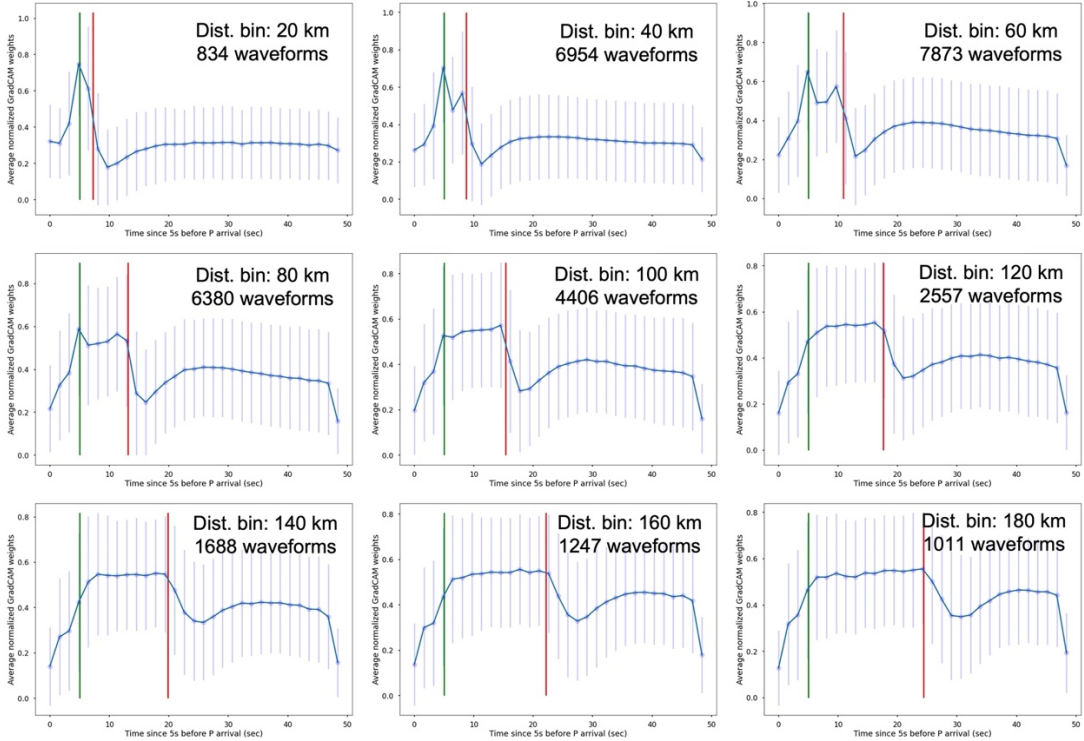

Figure S14. Average of the normalized Grad-CAM weights for the earthquake records across different distance bins on the time series input data. Distance bins and number of waveforms in each bin are shown in the titles. The blue thick lines are the average weights and the vertical thin blue lines are the standard deviation in the bins. The vertical green and red lines are the average of the estimated P and S arrivals using the same regional velocity models as in the calculation of the

P/S ratios. For each panel, the horizontal axis is the time in seconds starting 5s before the arrival of P wave. The vertical axis is the normalized weight.

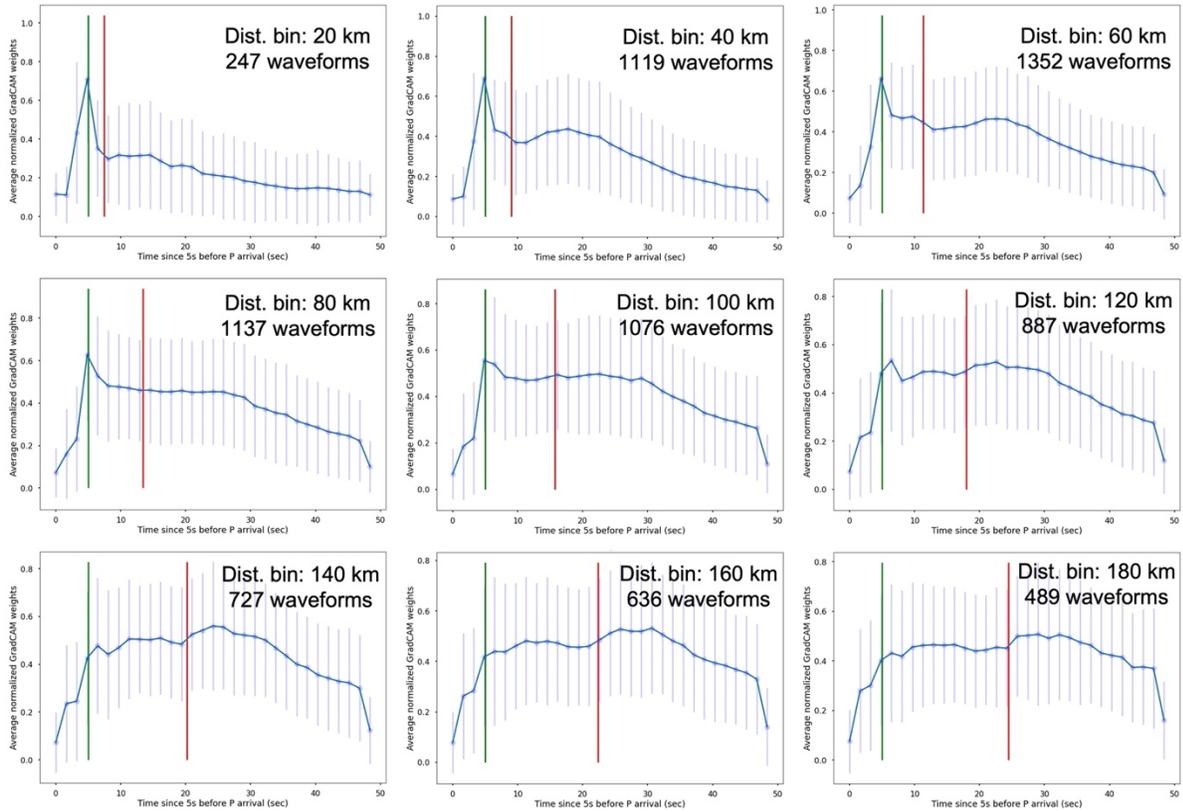

Figure S15. Average of the normalized Grad-CAM weights for the explosion records across different distance bins on the time series input data. Legends are the same as Figure S14.

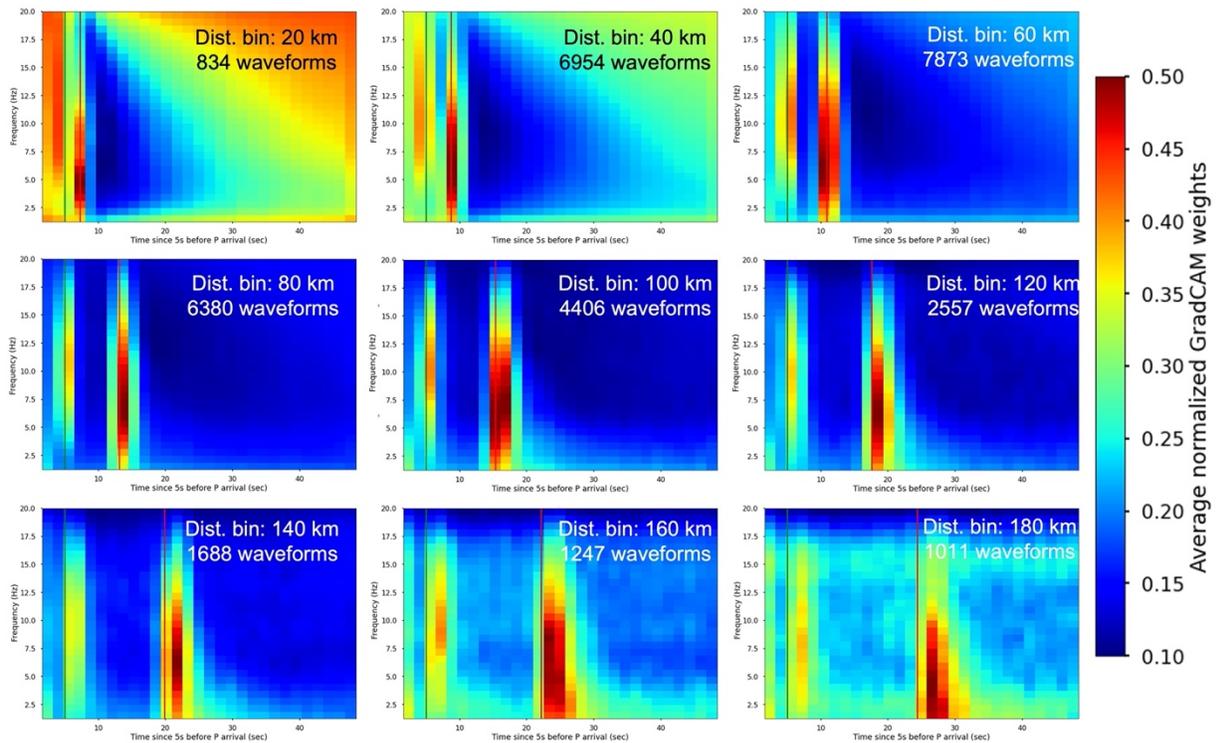

Figure S16. Average of the normalized Grad-CAM weights for the earthquake records across different distance bins on the spectrogram input data. Color shows the weights, x axis shows the time in seconds, and the y axis is the frequency in Hz. Distance bins and number of waveforms in the bins are shown in the titles. The vertical green and red lines are the average of the estimated P and S arrivals using the same regional velocity models as in the calculation of the P/S ratios.

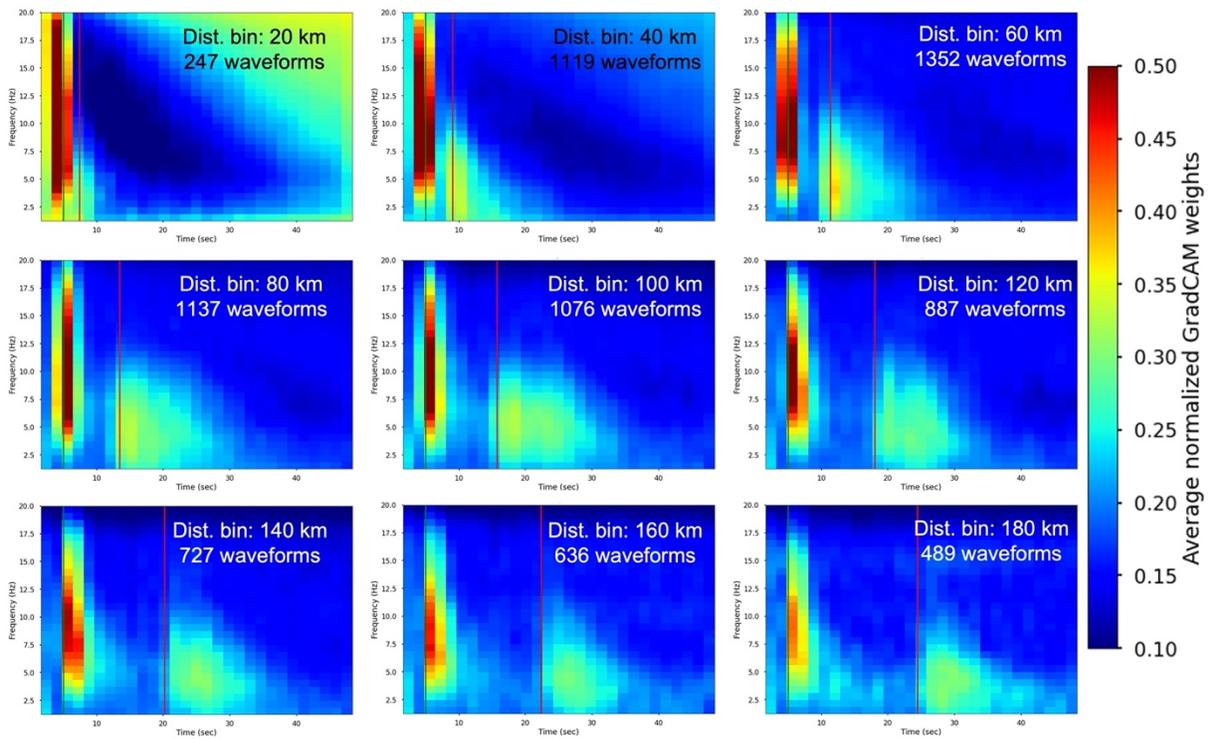

Figure S17. Average of the normalized Grad-CAM weights for the explosion records across different distance bins on the spectrogram input data. Legends are the same as Figure S16.

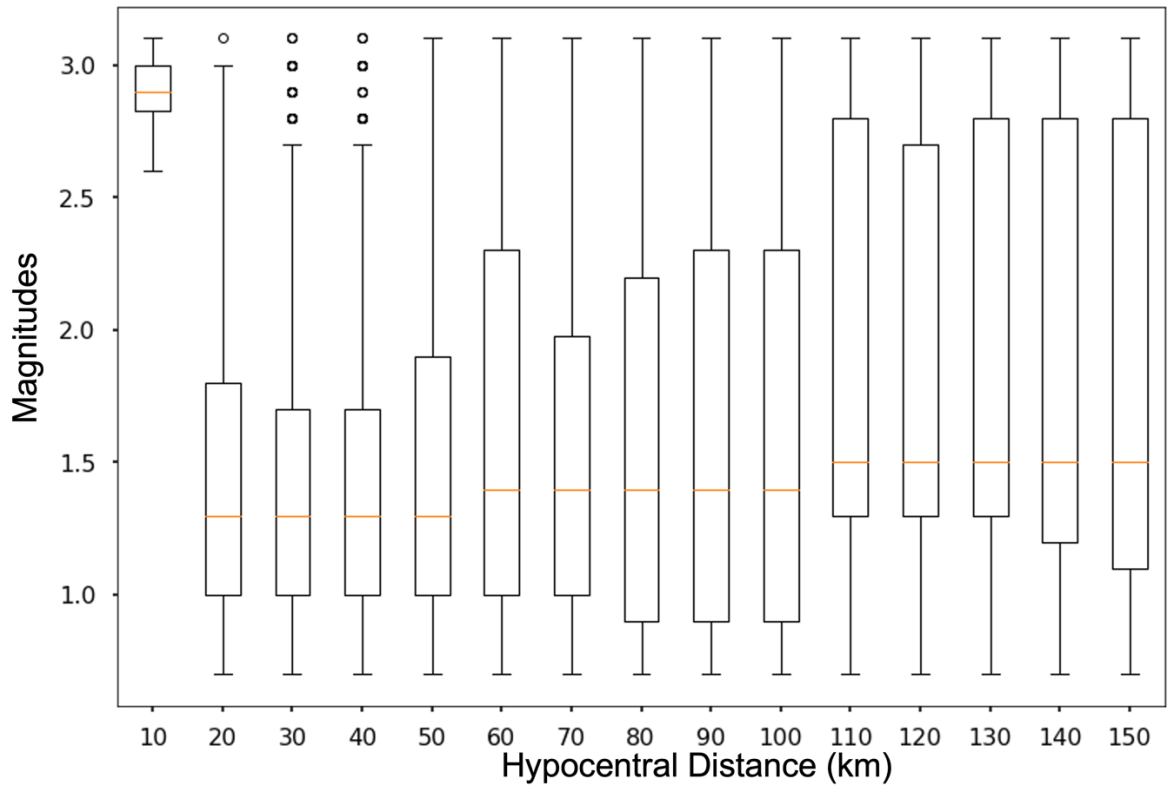

Figure S18. Boxplot of the magnitude distribution for the explosion data from the four regions used in training and testing dataset. Orange line shows the median value, and the top and bottom lines of the box indicate the first quartile (Q1) and the third quartile (Q3). The bars represent the whiskers extend from the box by 1.5x the inter-quartile range (IQR), i.e., Q1 – 1.5IQR and Q3+1.5IQR

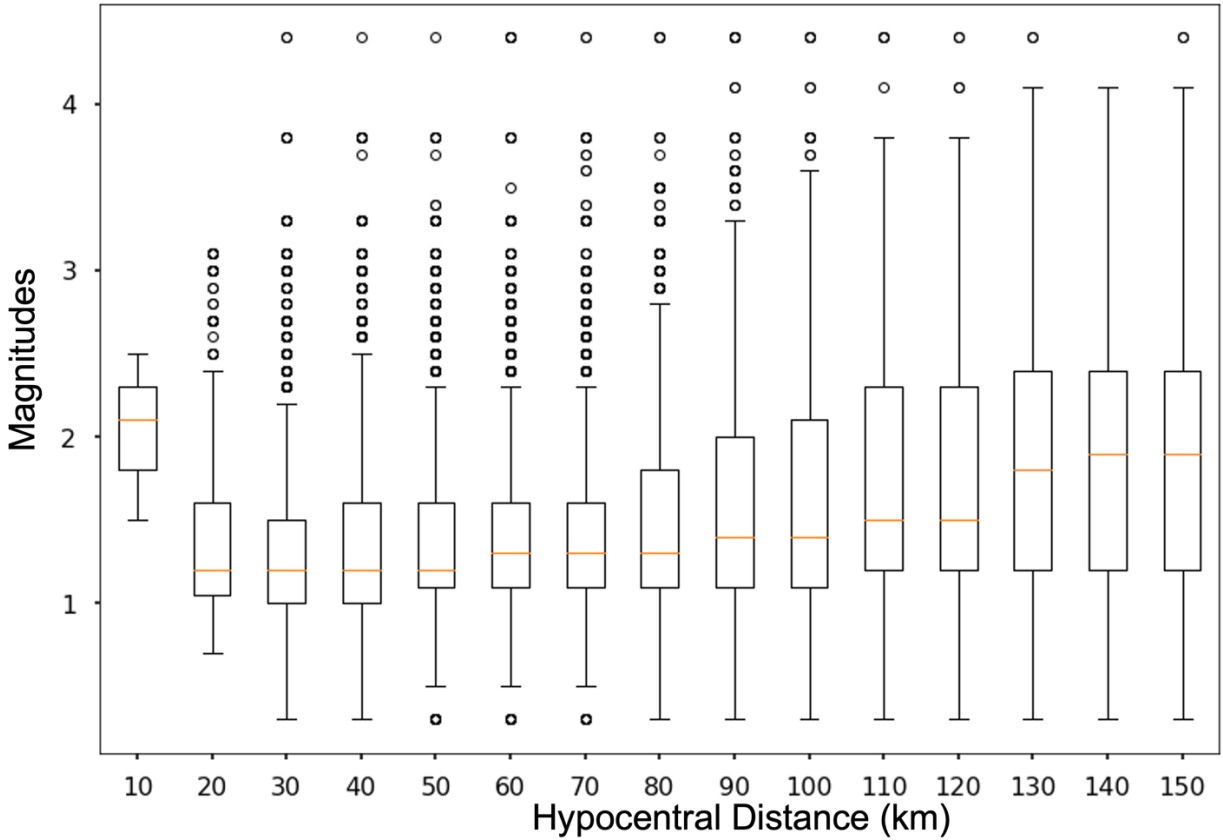

Figure S19. Boxplot of the magnitude distribution for the earthquake data from the four regions used in training and testing dataset. Orange line shows the median value, and the top and bottom lines of the box indicate the first quartile (Q1) and the third quartile (Q3). The bars represent the whiskers extend from the box by 1.5x the inter-quartile range (IQR), i.e., Q1 – 1.5IQR and Q3+1.5IQR